\documentclass[12pt]{article}

\usepackage{amssymb}
\usepackage{graphicx}
\usepackage{multicol}
\usepackage{vmargin}
\usepackage[latin1]{inputenc}
\usepackage{caption2}

\def\bsqr#1#2{\vrule height #2pt width #1pt}
\def\eop{\hfill\hbox{\bsqr 66}}

\newtheorem{lemma}{Lemma}

\newtheorem{prop}{Proposition}

\newcommand{\R}{\mathbb{R}}

\begin{document}

\title{A continuous stochastic model for cell sorting}

\author{Mathieu Emily \and Olivier Fran\c cois}

\maketitle

\begin{center}
{\small TIMC-TIMB, Faculty of Medicine, 38706 La Tronche cedex, France}
\end{center}

\begin{abstract}
\noindent
The differential Adhesion Hypothesis (DAH) is a theory of the organization of cells within a tissue. In this study we introduce a stochastic model supporting the DAH, that can be seen as a continuous version of a discrete model of Graner and Glazier. Our approach is based on the mathematical
framework of Gibbsian marked point processes. We provide a Markov chain
Monte Carlo algorithm that can reproduce classical biological patterns, and we propose an estimation procedure for a parameter that quantifies the strength of adhesion between cells. This procedure is tested through
simulations.
\end{abstract}

\noindent
{\it {\bf Keywords: } Cell-Cell Interactions - Gibbsian Model - Pseudo-likelihood - Tumorigenesis.}

\newpage

\section{Introduction}

The development and the maintenance of multi-cellular organisms are driven by permanent rearrangements of cell shapes and positions. Such rearrangements are a key step particularly towards reconstruction of functional organs \cite{Arm:89}. {\it In vitro} experiments such as Holtfreter's experiments for the pronephros \cite{Hol:43} and the famous example of an adult living organism Hydra \cite{Gie:72} are representative example of spectacular spontaneous sorting.

Among the pioneering works, Steinberg used the ability of cells to self-organize in coherent structures to conduct a series of experimental studies that characterized cell adhesion as a major actor of cell sorting \cite{Ste:62a,Ste:62b,Ste:62c,Ste:63}. Following his experiments, Steinberg suggested that the interaction between two cells involve an adhesion surface energy which varies according to the cell type. He interpreted cell sorting via the Differential Adhesion Hypothesis (DAH), which states that cells can explore various configurations and finally arrive to the lowest-energy configuration. During the past decades, the DAH has been experimentally tested in various situations such as gastrulation \cite{McC:87}, cell shaping \cite{Nub:90}, control of pattern formation \cite{New:90} and the engulfment of a tissue by another one. Recently, some of these experiments have been improved to evaluate the DAH rigorously \cite{Fot:05}.

Many mathematical models have been previously developed for the DAH. Most of these models are dedicated to the simulation of physical processes in agreement with the DAH: model simulations tend to minimize an Energy functional, called the Hamiltonian, supporting the DAH. Four groups of models can be distinguished according to the geometry of the tissues. First, {\em cell-lattice models} own the particularity that each cell is geometrically described by a common shape, generally a regular polygon (square, hexagon, etc ...) (see \cite{Moc:96} for example). The second class of models has been called {\em centric models}. This class provides more realistic cell geometries by using tesselations (Dirichlet tesselation for example) to define cell boundaries from a point pattern where points characterize cell centers \cite{Hon:96}. The third class of models is called the {\em vertex models}. These models are dual to the centric models \cite{Nag:01,Hon:04}. Thr fourth class of models, called {\em sub-cellular lattice models}, has been developed as a trade-off between the simulation speed of cell-lattice models and the geometrical flexibility of the centric models. These models have been introduced by Graner and Glazier \cite{Gra:92}, and they are also referred to as the {\em discrete GG model} afterwards.

In this study, a continuous model for cell sorting derived from the discrete GG model is presented. In this context, the geometry of cells is actually similar to the centric models: assuming that cell centers are known, the cells are approximated by Dirichlet cells. In agreement with the DAH, the new model is also based on an Hamiltonian inspired from the GG model Hamiltonian (see Section \ref{Sec:Model}). Following this approach, the model falls into a well-defined mathematical framework: Gibbsian marked point processes \cite{Van:00}. This mathematical background will allow us to better control the simulation procedure for generating cell sorting patterns.

Furthermore, recent developments in molecular biology emphasize the fact that cell-cell interactions play a major role in tumorigenesis \cite{Bar:00}. The nature of the interactions may actually reflect the initiation of a cancer. In addition, the invasive nature of a tumor is directly linked to the modification of the strength of cell-cell interactions \cite{Loz:03}. In this perspective, an important challenge is to quantify the strength of the adhesion between cells. Another goal of this study is therefore to provide an inference procedure for the parameter that governs the strength of
cell-cell adhesion.

The article is structured as follows. In Section \ref{Sec:Model}, the continuous model is introduced. Section \ref{Sec:Math} presents an estimator of the adhesion strength parameter, and gives some mathematical properties of the model. In Section \ref{Sec:Results}, results concerning simulations of cell sorting patterns
and the performance of the statistical estimator are presented.

\section{A continuous cell model for DAH}\label{Sec:Model}

This section introduces a new continuous model for differential adhesion. As in the previous approaches, the model is based on an Energy functional that describes cell-cell interactions. In the continuous model, the tissue is still described by a centric model where the points correspond to the locations of the cell centroids, and the marks correspond to the cell types. Honda's studies showed that the geometry of Dirichlet cells was in agreement with biological tissues for which the spatial coordinates of the nuclei were extracted thanks to molecular markers \cite{Hon:78,Hon:83}.

\paragraph{The GG model.}
Before describing our model, we start by giving a brief account on the GG model \cite{Gra:92}. In the GG model, a cell was not defined as a simple unit, but instead consisted of several pixels. The pixels could belong to three types: high surface energy cells, low surface energy cells or medium cells, which were coded as 1,2 and -1 respectively. According to the DAH, the Hamiltonian $H_{\rm GG}$ was defined as an extension of the Potts model as follows
\begin{equation}\label{Eq:Gra}
H_{\rm GG} = \sum_{(i,j) \sim (i^{\prime},j^{\prime})} J\left( { \tau (\sigma_{i,j}), \tau(\sigma_{ i^{\prime},j^{\prime}}) } \right) \left(1-\delta_{\sigma_{i,j},\sigma_{i^{\prime},j^{\prime}}} \right)
+ \lambda \sum_{\sigma} ( a(\sigma) - A_{\tau (\sigma)} )^2 \Gamma(A_{\tau (\sigma)}),
\end{equation}
where $(i,j)$ were the pixel spatial coordinates, $\sigma_{ij}$ represented the cell to which the pixel $(i,j)$ belonged, $\tau(\sigma_{ij})$ denoted the type of the cell $\sigma_{ij}$, and the function $J$ characterized the interaction intensity between two cell types ($\delta$ denoted the Kronecker symbol). In particular, the term $\left(1-\delta_{\sigma_{i,j},\sigma_{i^{\prime},j^{\prime}}}\right)$
indicated that the interaction between two pixels within the same cell was zero. Shape constraints were modeled by the second term where $\lambda$ corresponded to an elasticity coefficient, $a(\sigma)$ was the cell area and $A_{\tau (\sigma)}$ was a target area that depended on the cell type. The function $\Gamma$ denoted the Heavyside function. It was included in the formula so that medium cells (coding -1) were not subject to the shape constraint.

\paragraph{The continuous model.}
Here we denote by $(x_i)$ ($i= 1, \dots, n$) the $n$ cell centers. The Dirichlet cell of $x_i$ is denoted by Dir$(x_i)$, and is defined as the set of points which are closer to $x_i$ than to any other cell nucleus. In addition we write $|\rm{Dir}(x_i)|$ to denote the area of the cell Dir$(x_i)$. A continuous version of Equation \ref{Eq:Gra} can be constructed as follows. In the GG model, a cell $\sigma$ is in neighborhood of a cell $\sigma^{\prime}$ as soon as a single pixel of $\sigma$ is adjacent to a pixel from $\sigma^{\prime}$. With this in mind, the GG model's Hamiltonian can be rewritten as
$$
H_{\rm GG} = \sum_{\sigma \sim \sigma^{\prime}} |\sigma \cap \sigma^{\prime}| J\left( { \tau (\sigma), \tau(\sigma^{\prime}) } \right) + \lambda \sum_{\sigma} (a(\sigma)-A_{\tau (\sigma)})^2 \Gamma(A_{\tau (\sigma)})
$$
where $|\sigma \cap \sigma^{\prime}|$ is the number of connected pixels between $\sigma$ and $\sigma^{\prime}$, which can be identified as the Euclidian length of the interaction surface between the two cells $\sigma$ and $\sigma^{\prime}$. Identifying cells to their centers, $|\sigma \cap \sigma^{\prime}|$ can be approximated as $|{\rm Dir} (x_i \cap x_j)|$, where $|{\rm Dir} (x_i \cap x_j)|$ is the length of the Dirichlet edge shared by $x_i$ and $x_j$. In addition a cell area in our model matches with the area of a Dirichlet cell, which means that $a(\sigma)$ corresponds to $|{\rm Dir}(x_i)|$. Defining a (marked) cell configuration as
\begin{equation}\label{Eq:Conf}
\varphi = \{(x_1,\tau_1),\dots,(x_n,\tau_n)\},
\end{equation}
where the $(x_i)$ are the cell centers, and the $(\tau_i)$ are the corresponding cell types, we define here an equivalent Hamiltonian function
\begin{equation}
H(\varphi) = \sum_{i \sim j} |{\rm Dir}(x_i \cap x_j)|J(\tau_i,\tau_j) + \lambda
\sum_{i} (|{\rm Dir}(x_i)| - A_{\tau_i})^2 \Gamma(A_{\tau_i}).
\label{Eq:MyModel}
\end{equation}
The symbol $i \sim j$, means that the cells $x_i$ and $x_j$ share a common edge in the Dirichlet tiling. As in the GG model, this Hamiltonian can be viewed as the sum of two terms. The first term corresponds to pair potentials and controls the adhesion forces between contiguous cells.  The second term corresponds to singleton potentials, and is analogous to the one introduced in the GG model. It controls the shape of the cells as well as their density.

An additional parameter, denoted $\theta$, is also included in the model. It is called the {\em adhesion strength}, and allows us to define the Energy functional of our model as $\theta H(\varphi)$. Because $\lambda$ is a free parameter, we see that $\theta$ actually controls the relative intensity of pair interactions (adhesion between cells). This parameter is of particular interest because an important benefit of the continuous approach is to allow consistent statistical estimation procedures for this parameter.

\paragraph{Surface tensions.}
Biological tissue configurations are often interpreted in terms of surface tension parameters. For instance,
checkerboard patterns are usually associated with negative surface tensions, whereas cell sorting patterns
are associated with positive surface tensions. When two distinct cell types are considered, the surface
tension between cells with the distinct types can be defined as
$$
\gamma_{12} = J(\tau_1,\tau_2) - \frac{J(\tau_1,\tau_1) + J(\tau_2,\tau_2)}{2}.
$$
where $\tau_1$, $\tau_2$ are the cell types. In \cite{Gla:93}, surface tensions were denoted
as $\gamma_{ld}$. In the next section, the interaction parameter will not correspond to those used in the
GG model exactly. However, they will be fixed so that the surface tensions are of the same order.

\section{Model simulation and parameter estimation} \label{Sec:Math}

There are two important benefits of assuming a continuous model for cell sorting.
Following this approach we will be able to 1) provide mathematical conditions for
warranting the convergence of the model simulation algorithm (such controls were missing from the
original discrete approach), 2) propose statistical procedures for the inference of the interaction
strength. To reach these objectives, we shall integrate our model in the theory of
Gibbsian marked point processes which provides a general framework for simulation and
parameter estimation (see \cite{Van:00,Mol:03}).

\paragraph{Simulation.} Simulation from the continuous model can be performed according to the Metropolis-Hastings algorithm. At each iteration, the algorithm randomly chooses between two operations: either inserting or deleting a cell within a well-delineated region. Insertion and deletion of a cell in the configuration has been implemented using local changes as described in \cite{Wat:81} and \cite{Ber:94}. In both cases, the algorithm computes the Energy variation $\Delta_H$ after one operation is performed. Then it may accept the operation with probability $p = \min(1,\exp(-\theta \Delta_H))$ otherwise it is rejected. Convergence results for this continuous state space Markov chain will be established afterwards.

\paragraph{Pseudo-likelihood inference.} In this section we propose an inference procedure for estimating the parameter $\theta$. Estimating $\theta$ actually provides information about the strength of cell-cell
interactions and adhesion. Here we resort to a classical approximation in statistics: the pseudo-likelihood method, first introduced by Besag in the context of the analysis of dirty pictures \cite{Bes:75}. For any configuration $\varphi$, the pseudo-likelihood is defined as the product over all elements of $\varphi$ of the following conditional probabilities
$$
{\rm PL}(\theta) = \prod_{(x_i,\tau_i) \in \varphi} {\rm Prob}((x_i,\tau_i) | \varphi, \theta)
$$
In this formula, the conditional probability of observing $(x_i,\tau_i)$ at $x_i$ given the configuration
outside $x_i$ can be described as
$$
{\rm Prob}((x_i,\tau_i) | \varphi, \theta) = \frac{\exp(-\theta H_{\varphi}(x_i,\tau_i))}{\sum_{m \in M}
\exp(-\theta H_{\varphi}(x_i,m))}
$$
where $M$ corresponds to the set of the possible cell types (or marks), and where $H_{\varphi}(x_i,\tau_i)$
represents the contribution of the cell $x_i$ in the expression of the Hamiltonian $H(\varphi)$, i.e.,
$$
H_{\varphi}(x_i,\tau_i) = \sum_{j \sim i} |{\rm Dir}(x_i \cap x_j)|J(\tau_i,\tau_j)
+ \lambda  (|{\rm Dir}(x_i)| - A_{\tau_i})^2 \Gamma(A_{\tau_i}) .
$$

We also consider the logarithm of the pseudo-likelihood given by
\begin{equation}
{\rm L}(\theta) = - \sum_{(x_i,\tau_i) \in \varphi} \left( \theta H_{\varphi}(x_i,\tau_i) + \log \sum_{m \in M} \exp(-\theta H_{\varphi}(x_i,m))   \right).
\label{Eq:Pseudo}
\end{equation}

Maximizing Equation \ref{Eq:Pseudo} provides an estimator of $\theta$, namely
$$
\hat{\theta}(\varphi) = {\rm argmax}_{\theta} {\rm L}(\varphi,\theta),
$$
which can be achieved using standard techniques.

\paragraph{Gibbsian marked point processes.}

We now recall some basic results about Gibbsian marked point processes. Gibbsian models, according to the
statistical physics terminology, have been introduced and largely studied in \cite{Rue:69} or \cite{Pre:76}.
The Gibbsian category is analogous to the Markov point processes introduced in spatial statistics by Ripley
and Kelly \cite{Rip:77} and reviewed in great details by Van Lieshout \cite{Van:00}.

Here, we restrict ourselves to {\it marked point processes} that have a density $f$ with respect to the standard
Poisson process. A realization of such a process is called a configuration and is denoted as
$\varphi$. When $\varphi$ has exactly $n$ points, we can write
$$ \varphi =\{(x_1, \tau_1),\dots,(x_n,\tau_n)\}, $$
as in Equation \ref{Eq:Conf}. A cell-mark couple $(x_i, \tau_i)$ is called a point.
According to the {\it Hammersley-Clifford-Ripley-Kelly} Theorem
(see \cite{Rip:77}), the density of the process conditional to the number of points is of the
following form
$$
f_n (\varphi) = \frac{\exp(-\theta H(\varphi))}{Z} = \frac{h_n(\varphi)}{Z}
$$
where $H$ is the Hamiltonian of the system, $Z$ is the partition function. Using this formalism, the convergence of the Markov chain generated by the above algorithm can be studied. We focus on the Harris-Recurrence property, which means that the convergence of the chain is ensured for any initial configuration with a non-zero probability. The following result can be stated.

\bigskip

\begin{prop}\label{Pro:HR}
Let us consider a Gibbsian marked point process as defined in Equation \ref{Eq:MyModel}, and
$$
H(\varphi) = \sum_{i \sim j} |{\rm Dir}(x_i \cap x_j)|J(\tau_i,\tau_j) + \lambda \sum_{i} (|{\rm Dir}(x_i)| - A_{\tau_i})^2 \Gamma(A_{\tau_i}),
$$
where $J$ charaterizes the interaction intensity and $\lambda$ is an elasticity parameter.

Assume that $J$ is bounded \textmd{($|J| \leq K_J$)} and that two neighboring points are at a distance between $d$ and $D$ ($0 < d < D < +\infty$). Then the Markov Chain generated by the simulation algorithm is Harris-Recurrent.
\end{prop}

The proof of proposition \ref{Pro:HR} can be derived along the same lines as \cite{Gey:94} (Section~4, p.~364). It can be sketched as follows. First, it is clear that the transition probabilities of the proposed algorithm satisfy Equations 3.5-3.9 in \cite{Gey:94} (p.~361-362). Next, in order to ensure the irreductibility of the Markov chain, the density of the process have to be hereditary (Definition 3.1 in \cite{Gey:94}, p.~360), which is detailed just below. Then by adapting the proof of Corollary 2 in Tierney (\cite{Tie:94}, Section 3.1, p.~1713), it follows that the chain is Harris recurrent.

In our context, the herediraty property is a direct consequence of the local stability of the process. Local stability means that the Energy variation can be controled under local modifications of a configuration. The following result can be used.

\begin{lemma}[local stability]\label{Lem:Stab}
Under the same conditions as in proposition \ref{Pro:HR}, there exists a constant $C(K_J,\lambda,d,D)$ so that
$$
| H(\varphi \cup (x_{n+1},\tau_{n+1}))  - H(\varphi)| \leq C(K_J,\lambda,d,D)
$$
where the inequality holds for all configurations and all $x_{n+1} \in \R^2$ with type $\tau_{n+1}$. More specifically, the constant is equal to
$$
C(K_J,\lambda,d,D) =  \frac{N(N+1)}{2} \frac{D}{\sin (\beta)} K_J + \lambda (N+1) \pi^2 \left(\frac{D}{2\sin(\beta)} \right)^4
$$
where $N = 2 \pi / \beta$ the maximum number of neighbors of any cell, and $\beta$ is the minimum angle of all Delaunay's triangles.
\end{lemma}

\bigskip

The proof of Lemma \ref{Lem:Stab}  is given in the Appendix. The restriction to $J$ bounded and the other minor assumptions are necessary for technical reasons, and they may appear unrealistic as biological patterns are concerned. Nevertheless the situations encountered in the remainder of the article will always check these conditions.

\section{Results} \label{Sec:Results}

\subsection{Simulation of biological patterns} \label{Sec:Simu}

In this section, we report simulation results obtained with a finite set of marks $M=\{\tau_1,\tau_2,\tau_E\}$. As the GG model does, we show that the model has the ability to reproduce at least three kinds of biological relevant patterns: Checkerboard, Cell Sorting and Engulfment. Checkerboard pattern formation was investigated in a simulation study of Honda~{\it et al.}~\cite{Hon:86} about the sexual maturation of the avian oviduct epithelium. Cell sorting is a standard pattern of mixed heterotypic aggregates. Experimental observations of this phenomena were reported by Takeuchi {\it et al.} \cite{Tak:88} and Armstrong \cite{Arm:89}. Engulfment of a tissue by another one was studied by Armstrong \cite{Arm:89} and Foty {\it et al.} \cite{Fot:96}.  This phenomenon is a direct consequence of adhesion processes between the two cell types and the extracellular medium.

\begin{figure}
\centerline{
\includegraphics[height=7cm]{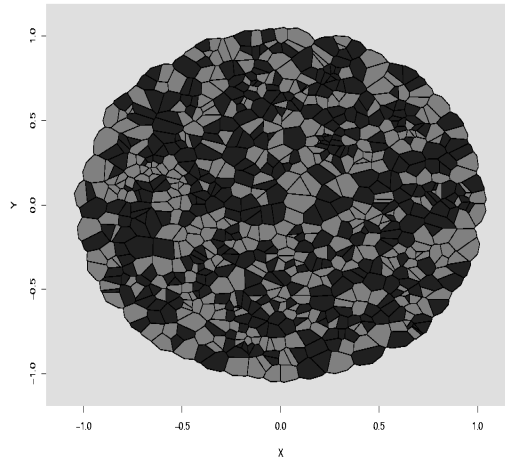}
}
\caption{\small{The initial configuration for simulating Checkerboard, Cell Sorting and Engulfment
patterns. It consists of about 1000 active cells surrounded by an extracellular medium. The active cells are randomly located in the unit sphere, and their types are randomly sampled from $M$. Cells of type $\tau_1$ are colored grey while cells of type $\tau_2$ are colored in black.}}
\label{Fig:Init}
\end{figure}

The two marks $\tau_1$ and $\tau_2$ represent ``active cell types" with distinct phenotypes responsible for the adhesion process. In addition, active cells are surrounded by an extracellular medium modeled by cells of type $\tau_E$. These three types are similar to the $l$, $d$ and $M$ types of Glazier and Graner \cite{Gla:93}. Simulations were generated from the Metropolis algorithm presented in the previous section. A unique configuration was used to initialize all the simulations. It consisted of about $1000$ randomly located active cells, and the configuration displayed in Figure \ref{Fig:Init}. In this configuration, the marks were also random. The target areas for active cells were equal to $A_{\tau_1} = A_{\tau_2} = 5.10^{-3}$. Under equilibrium, configurations were expected to consist of about $\pi / 5.10^{-3} \approx 628$ cells. No area constraint affected the $\tau_E$ cells and we set $A_{\tau_E} = -1$. The adhesion strength parameter $\theta$ was fixed to $\theta = 10$, and we adjusted the interaction intensities so that the surface tensions were of same order as those considered by Granier and Glazier \cite{Gla:93}.

\begin{figure}
\centerline{
\begin{tabular}{cp{0.5cm}c}
{\includegraphics[height=7cm]{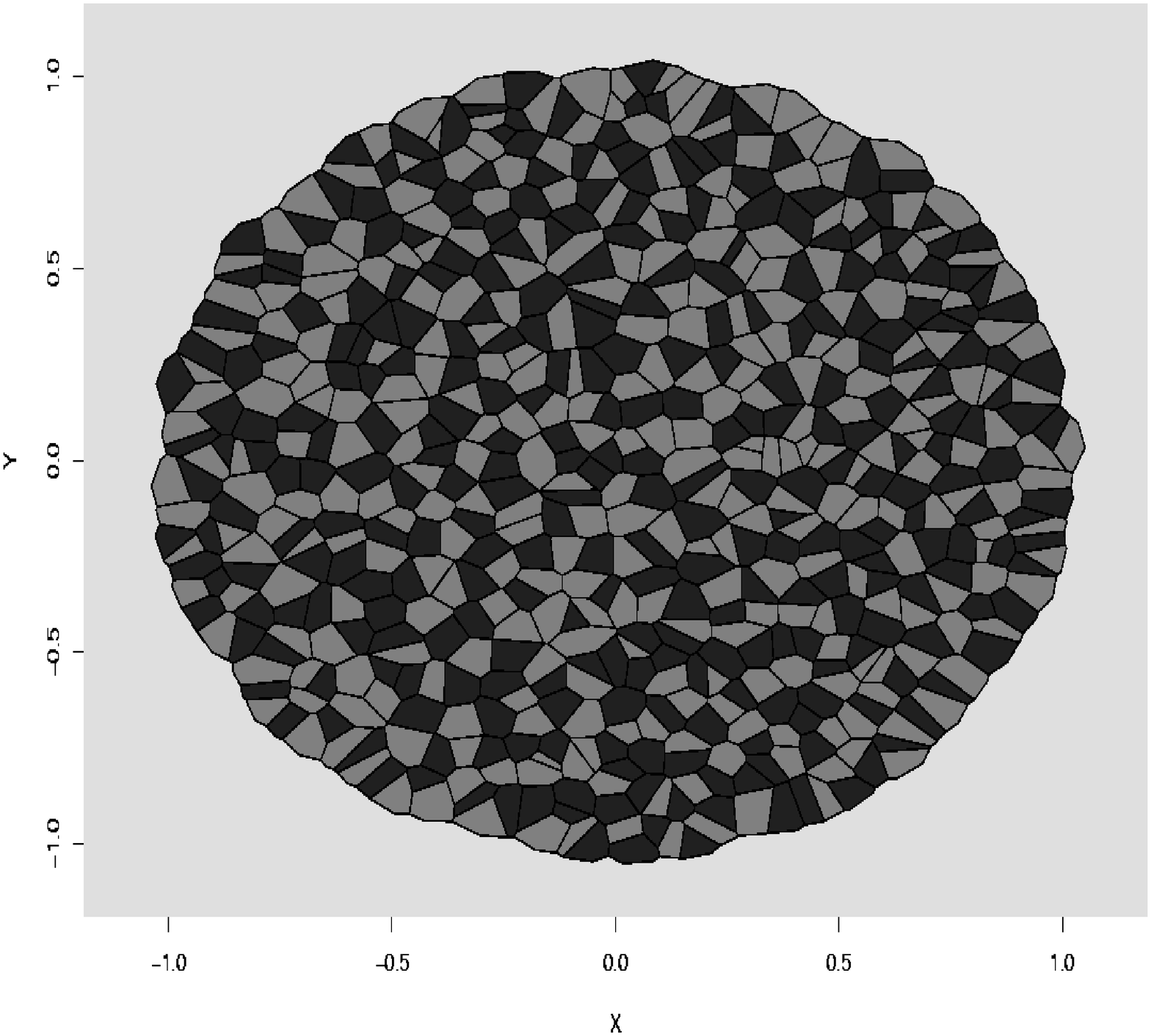}}&&
{\includegraphics[height=7cm]{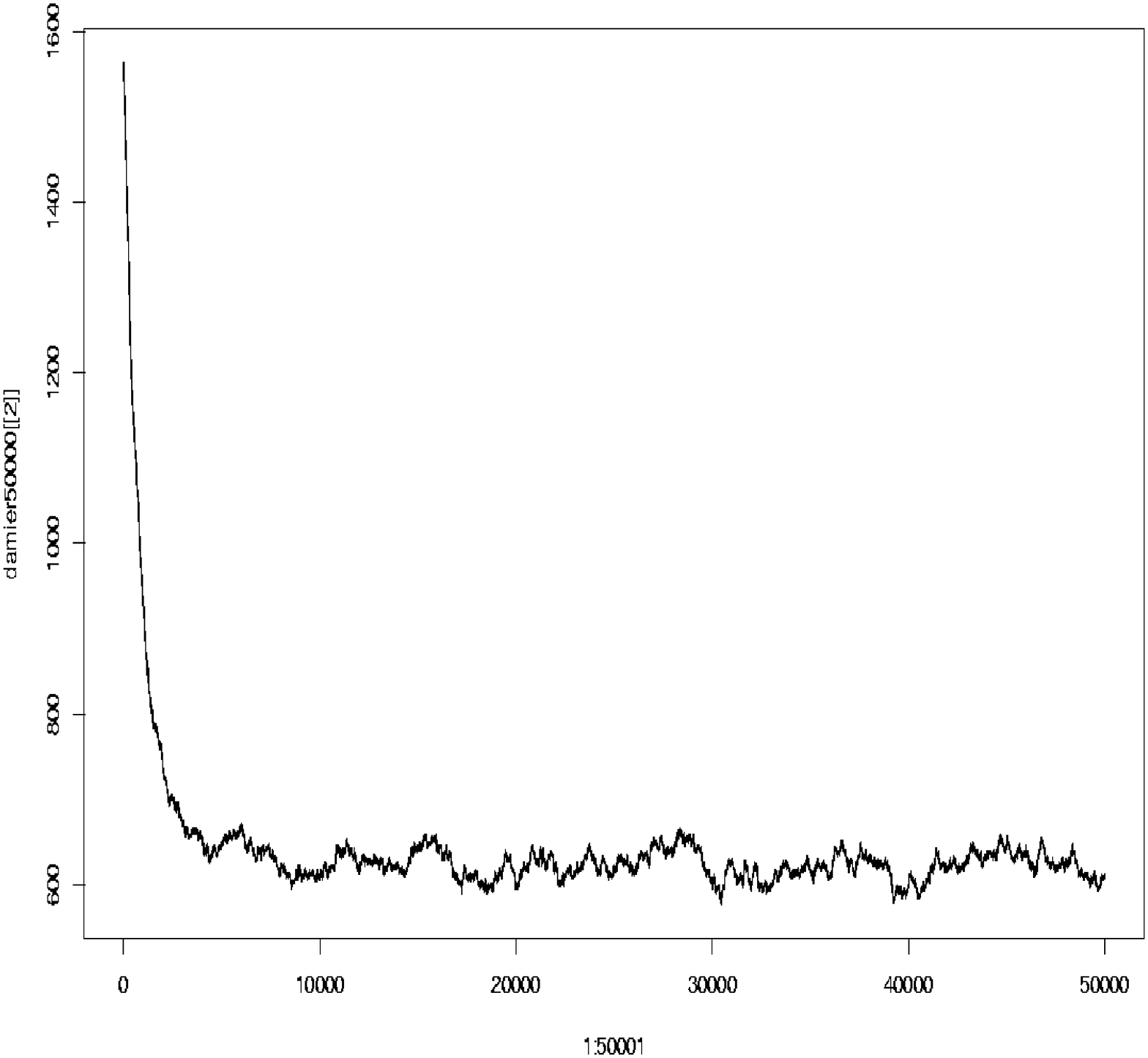}}\\
(a) && (b)
\end{tabular}
}
\caption{\small{Checkerboard simulation. The interaction intensities were chosen as follows: $J(\tau_1,\tau_1) = 1$, $J(\tau_2,\tau_2) = 1$, $J(\tau_1,\tau_2) = 0$, $J(\tau_1,\tau_E) = 0$ and $J(\tau_2,\tau_E) = 0$.  (a) The configuration obtained  after 50,000 iterations. (b) The decrease of the Energy as a function of the iteration steps.}}
\label{Fig:Checkerboard}
\end{figure}

Checkerboard patterns can be interpreted as arising from negative surface tensions. In the GG model, Checkerboard patterns were generated using parameter settings that corresponded to surface tensions around $\gamma_{12} = -3$. Figure \ref{Fig:Checkerboard} displays the configuration obtained after 50,000 cycles of the Metropolis-Hastings algorithm, where the interaction intensities were fixed at  $J(\tau_1,\tau_2) = 0$, $J(\tau_1,\tau_1) = J(\tau_2,\tau_2) = 1$ and $J(\tau_E,\tau_1) = J(\tau_E,\tau_2) = 0$. These interaction intensities corresponded to a surface tension equal to  $\gamma_{12} = -1$ which was of the same order as the one used in the GG model.

\begin{figure}
\centerline{
\begin{tabular}{cp{0.5cm}c}
{\includegraphics[height=7cm]{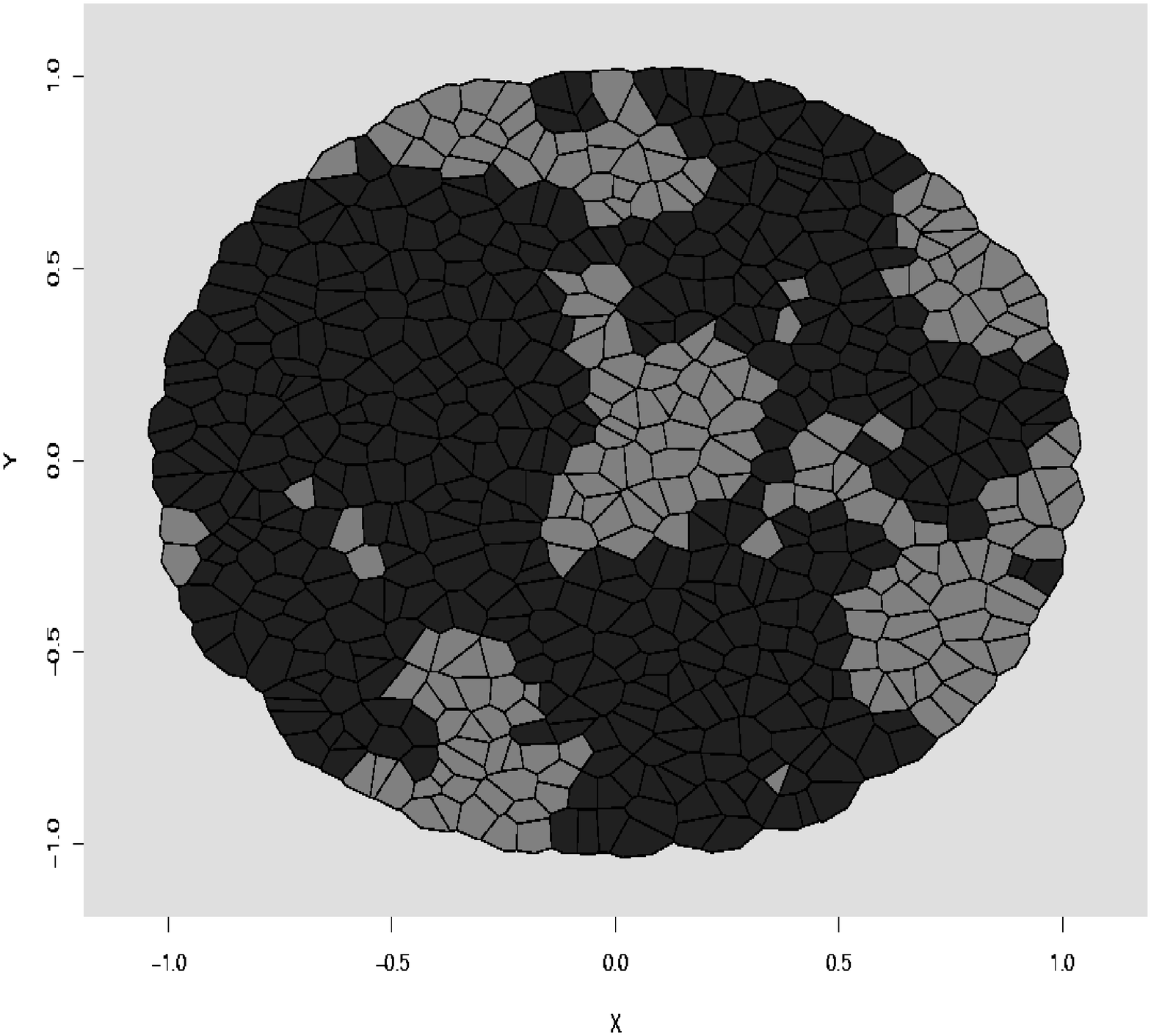}}&&
{\includegraphics[height=7cm]{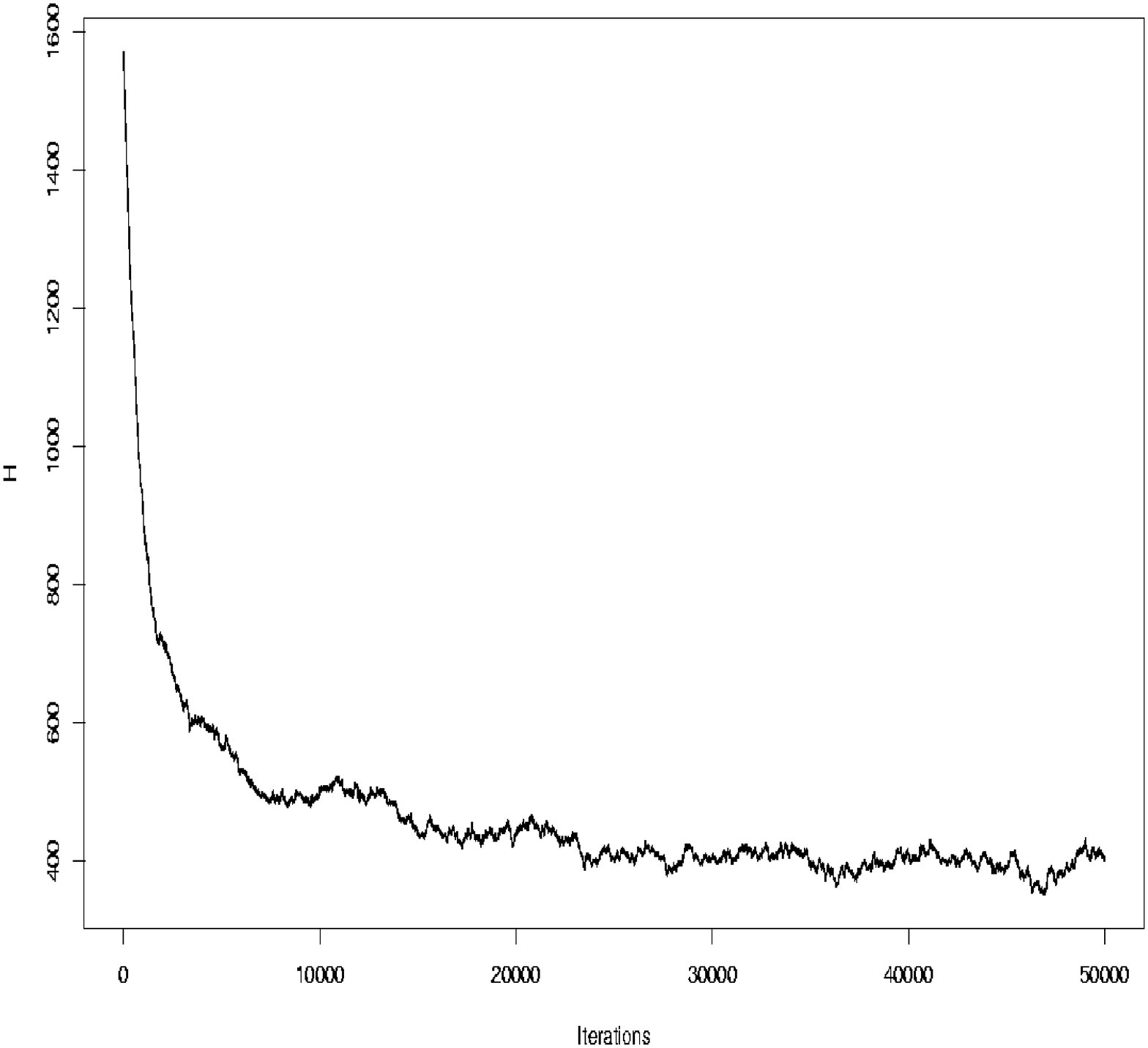}}\\
(a) && (b)
\end{tabular}
}
\caption{\small{Cell Sorting simulation. The interaction intensities were chosen as follows: $J(\tau_1,\tau_1) = 0$, $J(\tau_2,\tau_2) = 0$, $J(\tau_1,\tau_2) = 1$, $J(\tau_1,\tau_E) = 0$ and $J(\tau_2,\tau_E) = 0$.  (a) The configuration obtained  after 50,000 iterations. (b) The decrease of the Energy as a function of the iteration steps.}}
\label{Fig:CellSorting}
\end{figure}

In contrast, Cell Sorting patterns arise from positive surface tensions between active cells. In the GG model, Cell Sorting patterns were generated using parameter settings that corresponded to surface tensions around $\gamma_{12} = +3$. In our model, simulations were conducted using the following interaction intensities: $J(\tau_1,\tau_2) = 1$, $J(\tau_1,\tau_1) = J(\tau_2,\tau_2) = 0$ and $J(\tau_E,\tau_1) = J(\tau_E,\tau_2) = 0$. This corresponded to the value $\gamma_{12} = +1$. The configuration obtained after 50,000 steps is displayed in Figure \ref{Fig:CellSorting}.

\begin{figure}[hp]
\centerline{
\begin{tabular}{cp{0.5cm}c}
{\includegraphics[height=7cm]{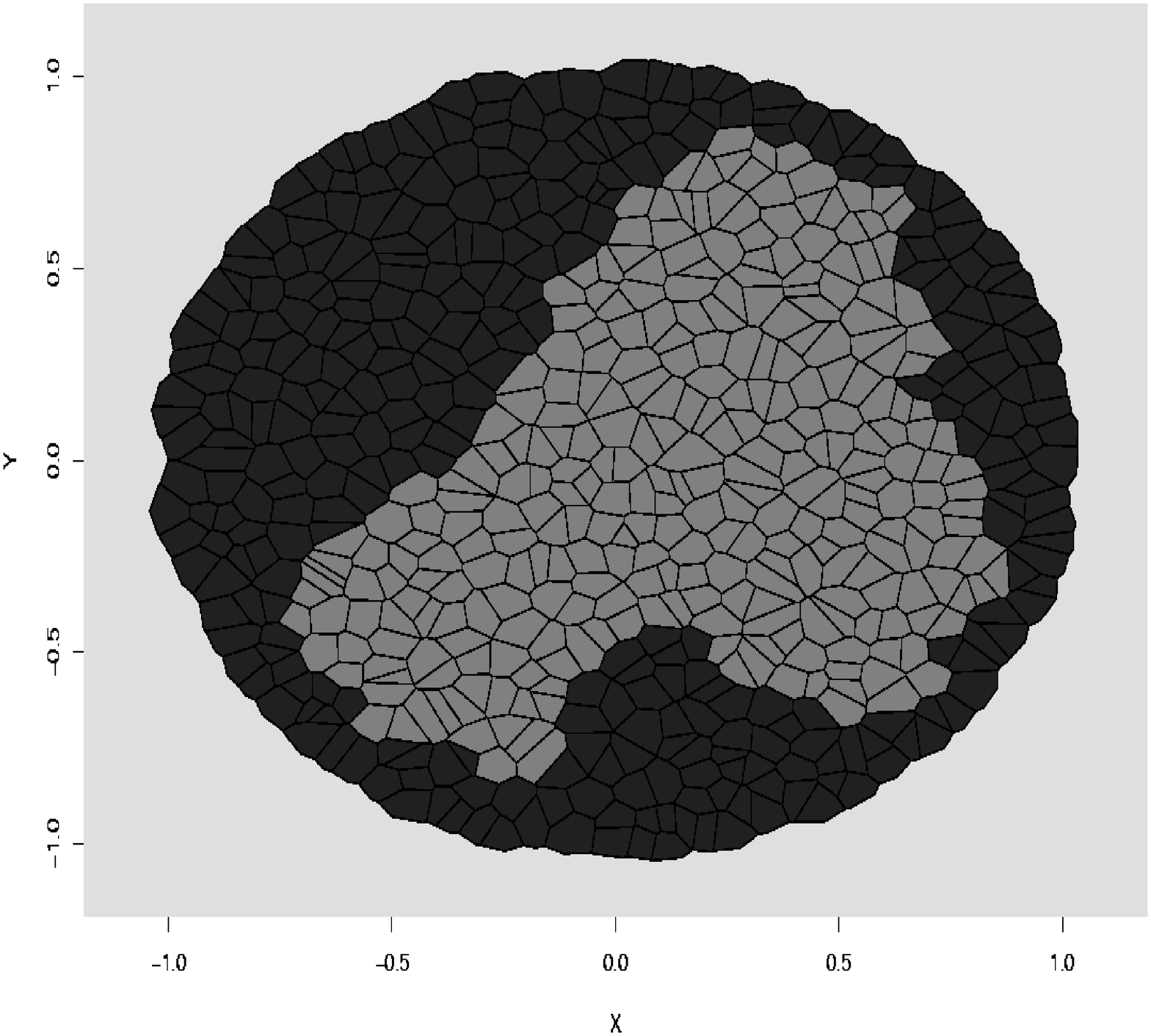}}&&
{\includegraphics[height=7cm]{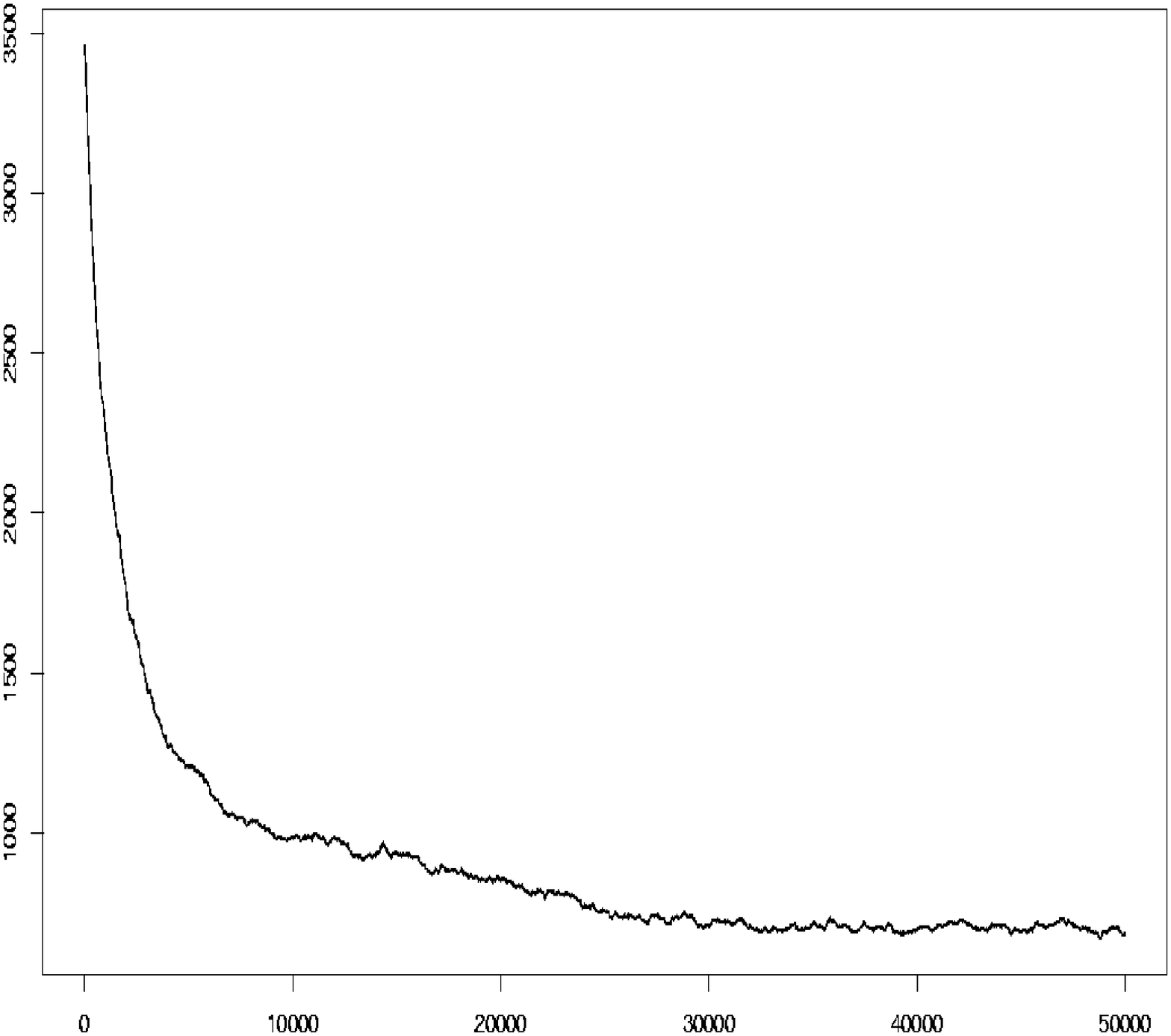}}
\end{tabular}
}
\caption{\small{Engulfment simulation. The interaction intensities were chosen as follows: $J(\tau_1,\tau_1) = 0$, $J(\tau_2,\tau_2) = 0$, $J(\tau_1,\tau_2) = 1$, $J(\tau_1,\tau_E) = 0$ and $J(\tau_2,\tau_E) = 1$.  (a) The configuration obtained  after 50,000 iterations. (b) The decrease of the Energy as a function of the iteration steps.}}
\label{Fig:Engulfment}
\end{figure}

Simulations of Engulfment were conducted using the following parameters: \\$J(\tau_1,\tau_2) = 1,\; J(\tau_1,\tau_1) = J(\tau_2,\tau_2) = 0,\; J(\tau_E,\tau_1) = 0,\; J(\tau_E,\tau_2) = 1$. These interaction intensities provided  positive surface tensions between active cells, which then tended to form clusters. The fact that $J(\tau_E,\tau_2)$ was greater than $J(\tau_E,\tau_1)$ ensured that $\tau_1$ cells were more likely to be close to the extracellular medium and to surround the $\tau_2$ cells. The results are displayed in Figure \ref{Fig:Engulfment}.

\subsection{Statistical estimation of the adhesion strength parameter}
\label{Sec:thetainfluence}

In this section, we study the influence of varying the adhesion strength parameter $\theta$ on simulation results, and we summarize the performances of the maximum pseudo-likelihood estimator $\hat \theta$.

To assess the influence of $\theta$ on pattern simulations, three values were tested: $\theta = 1$, $\theta = 5$ and $\theta=10$. The results are presented for simulations of Checkerboard and Cell Sorting patterns. In both cases, the interaction intensities were setting as in section \ref{Sec:Simu}, and provided $\gamma_{12}=-1$ in Checkerboard simulations and $\gamma_{12} = +1$ in Cell Sorting simulations.

\begin{figure}
\centerline{
\begin{tabular}{cp{0.5cm}cp{0.5cm}c}
\multicolumn{5}{c}{Checkerboard}\\
{\includegraphics[height=5cm]{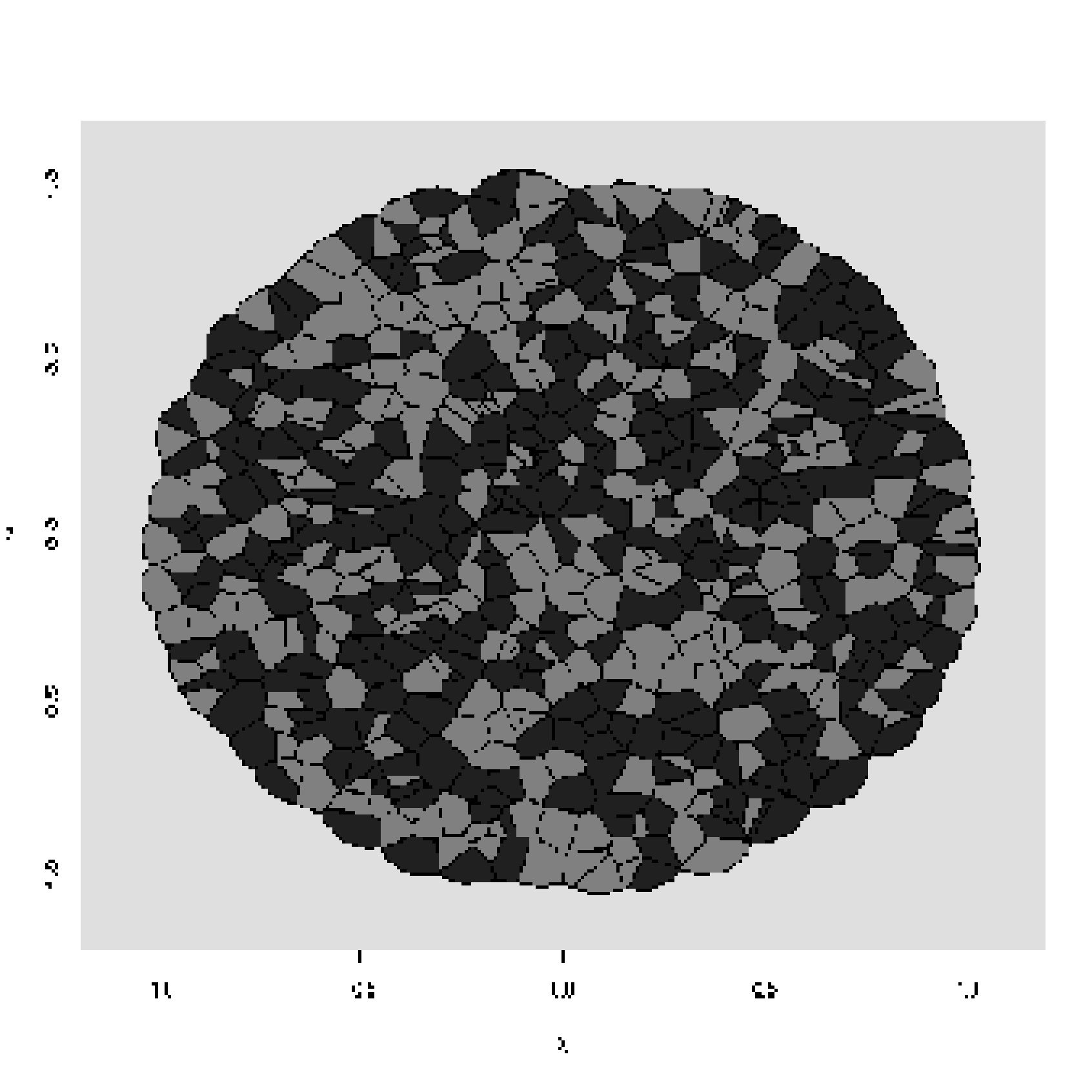}}&&
{\includegraphics[height=5cm]{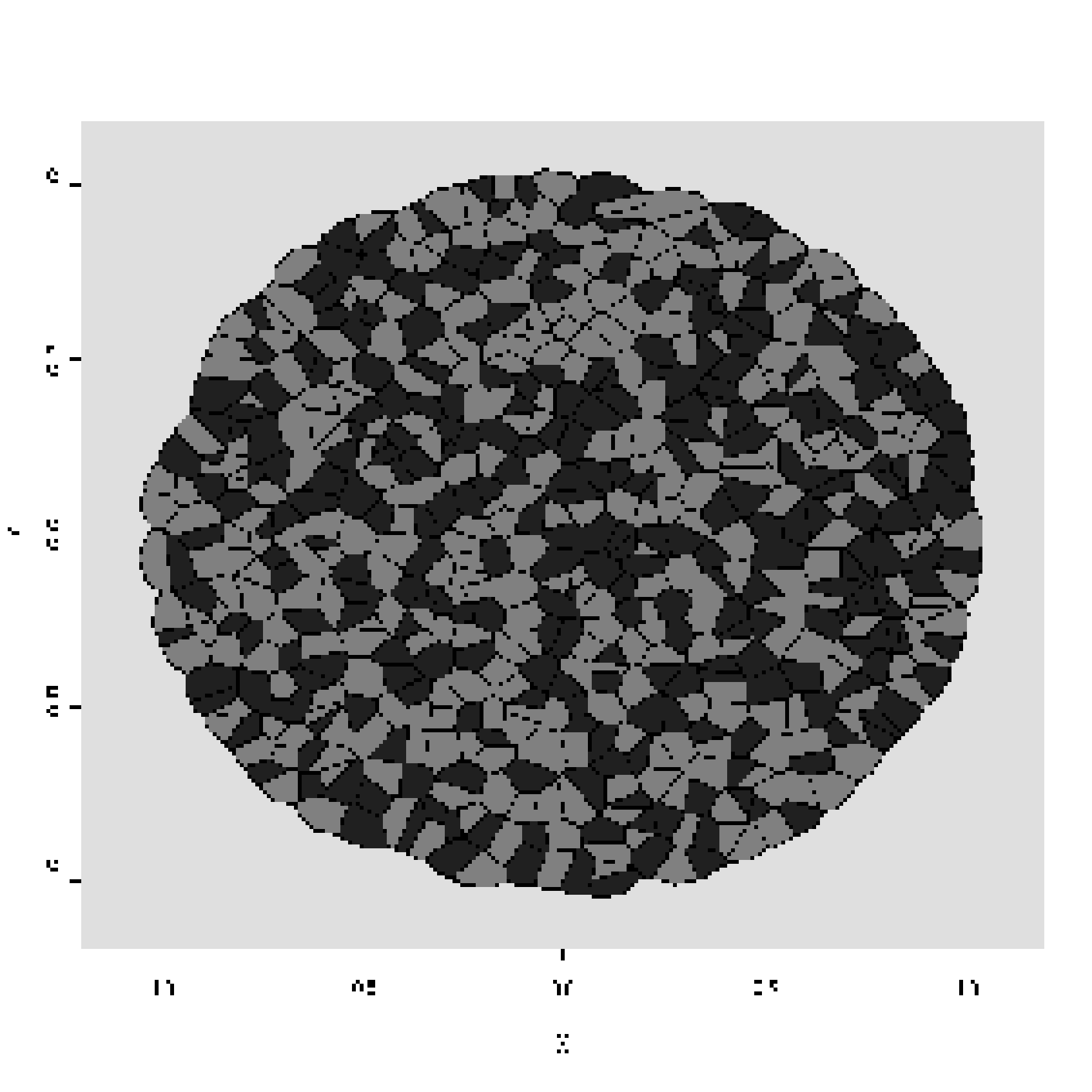}}&&
{\includegraphics[height=5cm]{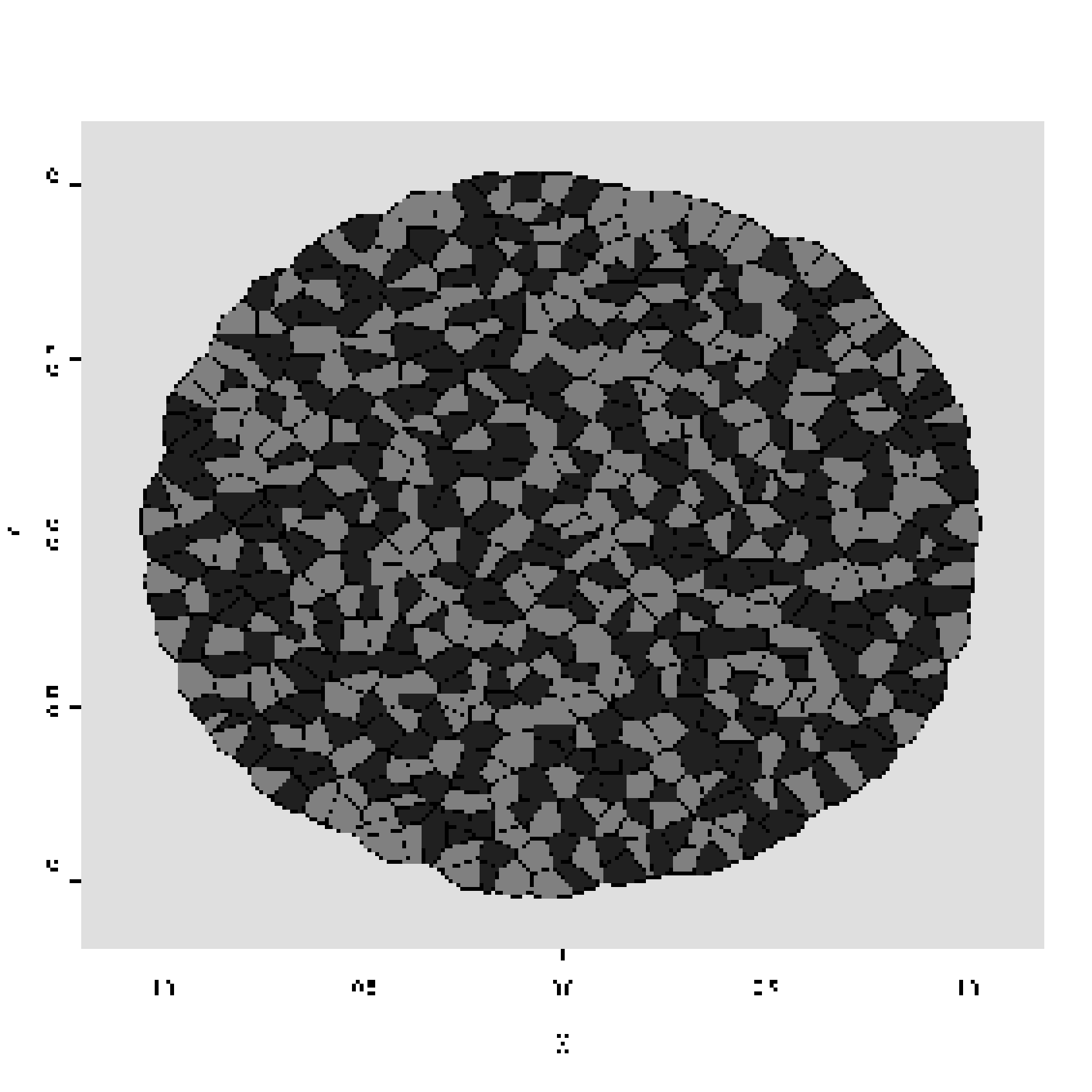}}\\
$\theta = 1$ && $\theta = 5$ && $\theta = 10$\\
\multicolumn{5}{c}{Cell Sorting}\\
{\includegraphics[height=5cm]{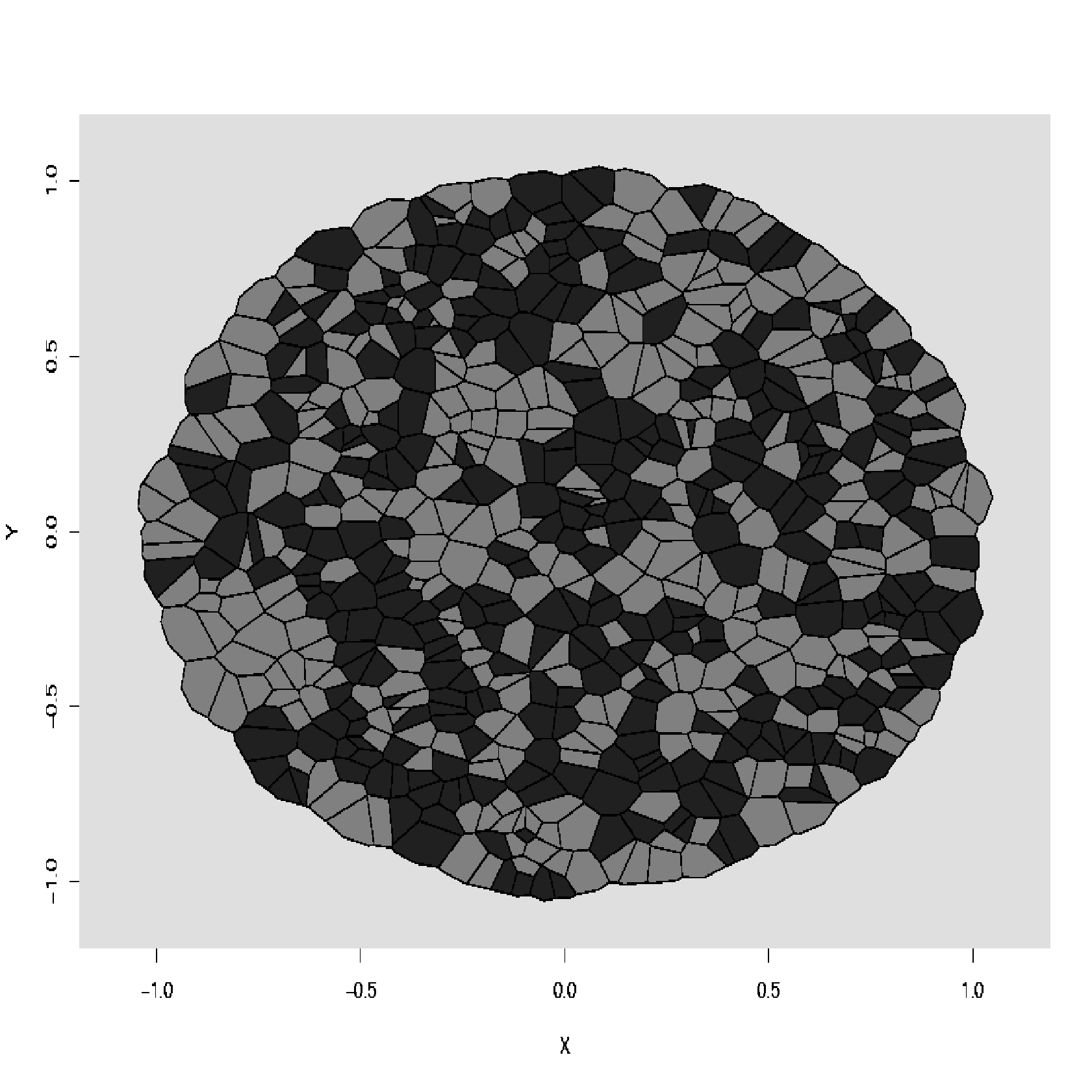}}&&
{\includegraphics[height=5cm]{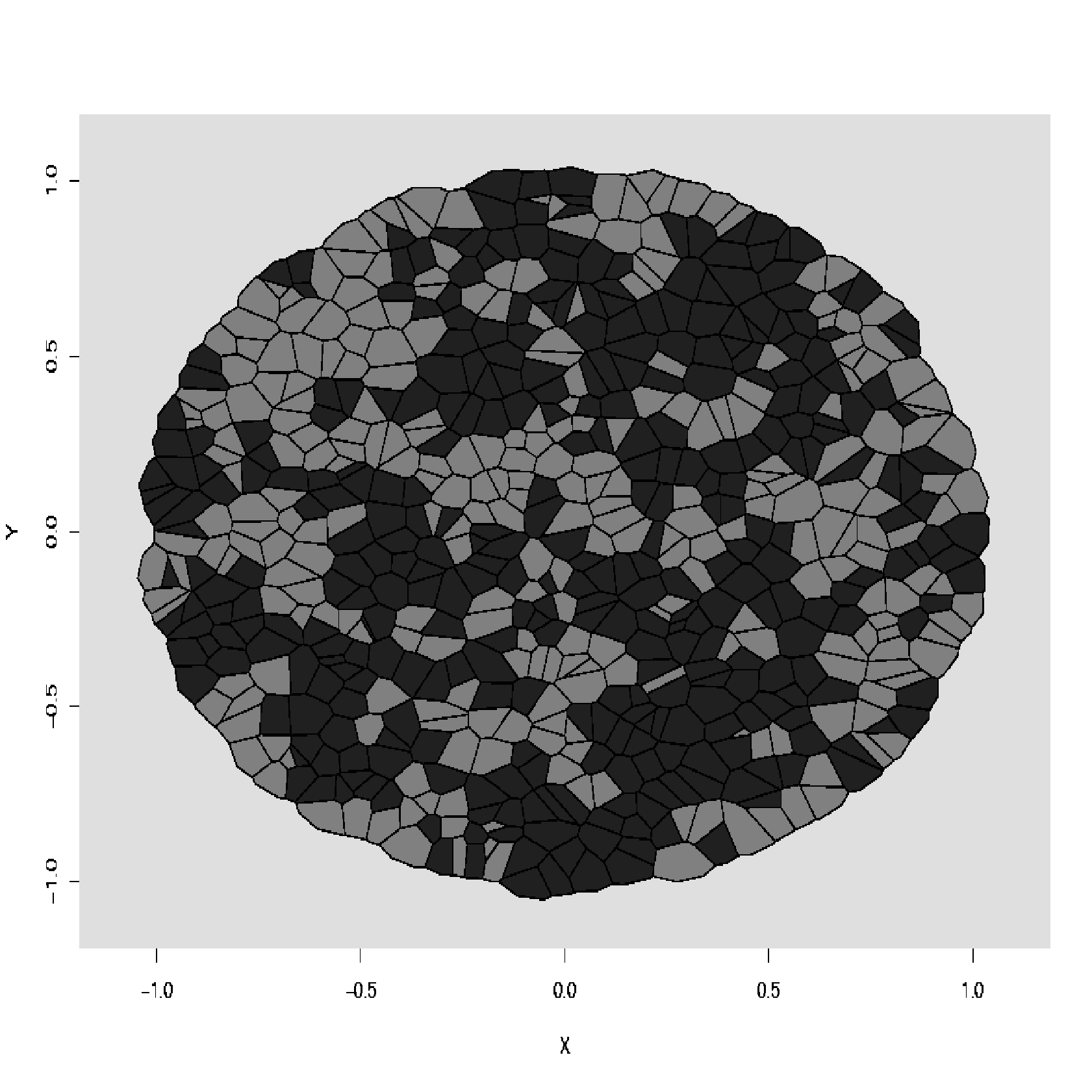}}&&
{\includegraphics[height=5cm]{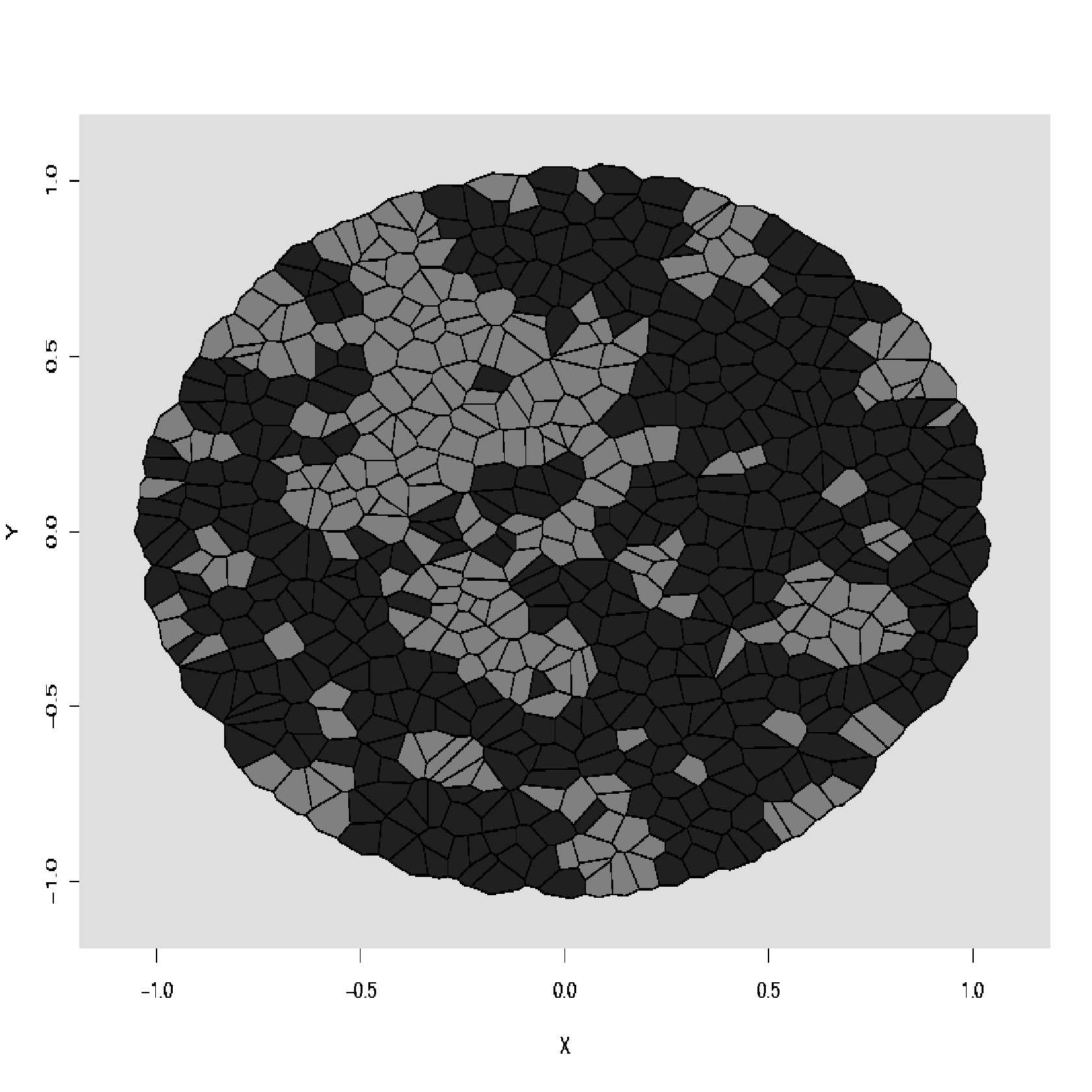}}\\
$\theta = 1$ && $\theta = 5$ && $\theta = 10$
\end{tabular}
}
\caption{\small{Influence of $\theta$ in Checkerboard simulations. Final configurations using three different values for $\theta$. Simulations gradually corresponds to either a Checkerboard or large clusters.}}
\label{Fig:TheChe}
\end{figure}

In all cases we ran the Metropolis algorithm for 50,000 steps, which were enough to provide a flat profile of Energy variation. The three final configurations, in both Checkerboard and Cell Sorting, are displayed in Figure \ref{Fig:TheChe}. Either for Checkerboard or for Cell Sorting simulations, we can see a gradual evolution as $\theta$ increases. For $\theta=1$, the marks are randomly distributed, for $\theta =5$ a small inhibition is visible in the Checkercoard simulation while small clusters appear in the Cell Sorting pattern. Finally, for $\theta=10$ the stronger inhibition between cells with the same types provides a more pronounced checkerboard pattern, and larger clusters are obtained in Cell Sorting.

We studied the statistical performances of $\hat{\theta}$ through simulations of Checkerboard and Cell Sorting patterns. The interaction intensities were the same as previously. For each value of $\theta$, 50 replicates were generated from which the mean and the variance of $\hat{\theta}$ was estimated. Table \ref{Tab:Estim} summarizes the results obtained for $\theta$ in the range ($1\dots 20$). For Cell Sorting, the bias was weak for all values of $\theta$, while for Checkerboard the bias seemed to be
slightly higher. The results were similar regarding the variance. It was higher for Checkerboard than for Cell Sorting. In addition the variance increased as $\theta$ increased.

\subsection{Real data}

\begin{figure}
\centerline{
{\includegraphics[height=6cm]{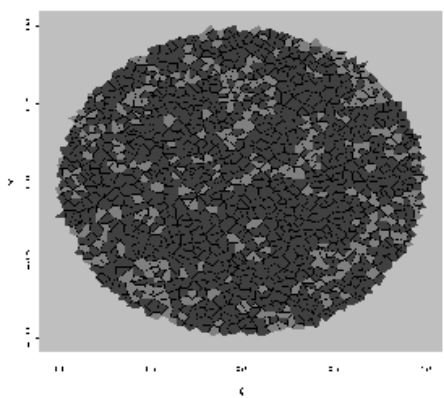}}
}
\caption{Real data. The statistical procedure was conducted using the interactions parameters for cell sorting patterns: $J(\tau_1,\tau_1) = 1$, $J(\tau_2,\tau_2) = 1$, $J(\tau_1,\tau_2) = 0$, $J(\tau_1,\tau_E) = 0$ and $J(\tau_2,\tau_E) = 0$. We obtained $\hat{\theta} \approx 4.79$.}
\label{Fig:real}
\end{figure}

Estimation of the adhesion strength was also performed on real data. We used survivin and beta-catenin markers in the context of medulloblastoma \cite{Piz:05}. These markers are known to be implicated in complexes that regulate adhesion between contiguous cells. An image analysis, analogous to the analysis performed in \cite{Emi:05}, has been achieved to extract the location of each cell nuclei and the level of expression of markers in each cell. The modeled tissue is displayed in Figure \ref{Fig:real}.

We considered that the data were relevant to a Cell Sorting pattern, and the corresponding interaction parameters were then used. The estimate of $\theta$ was computed as $\hat{\theta} \approx 4.79$ which fall into the range of well-fitted values.

\section{Discussion}

In this paper, a model based on marked point processes theory and inspired from \cite{Gra:92} has been studied. The new model proposes a continuous extension of the GG model by considering continuous instead of discrete cells, and it overcomes some limitations of the discrete model. First, in the discrete simulation algorithm cells are not constrained to be simply connected (i.e., they may be divided in non-connected components). Graner and Glazier solved this problem by choosing appropriate initial configurations and small temperatures.
Another issue related to the discrete model is that the discretization scale influences the choice of the parameters $\lambda$ and $T$ which in turn contribute to the acceptance/rejection probabilities. Finally the discrete algorithm was lacking good convergence properties. More specifically, the Markov chain generated by the discrete algorithm might not be ergodic and might also be influenced by the discretization scale. Again, these drawbacks did not impact the original results, because the authors chose their initial configurations so that the desired output was produced.

The algorithm proposed in section \ref{Sec:Math} provides convergent simulations whatever the initial configuration and is also considerably faster. In this study, we have shown that the new model was able to reproduce biologically relevant cell patterns such as Checkerboard, Cell Sorting and Engulfment. Furthermore, the model has been built so that it include the strength of cell-cell adhesion. We have proposed and validated an inference procedure based on the pseudo-likelihood. The statistical errors were low in Cell Sorting simulations. In Checkerboard simulations, bias and variance were slighlty higher than for Cell Sorting but still reasonable in the range of tested parameters.

Further improvements of this approach would require a deeper mathematical study which is beyond the scope of this study. In particular, the theory of marked point processes makes it possible to establish theoretical consistency results for $\hat{\theta}$. Furthermore, the other interaction parameters can also be estimated in the same way as $\theta$ was. Although we did not assessed the performances of these estimators, we believe that they would be useful for analyzing tissues arrays, as generated by high-throughtput cancer studies \cite{Kon:98}.

\bibliographystyle{empty}

\newpage
\section*{Appendix}

In this section, we give a proof of Lemma \ref{Lem:Stab}. This proof needs some geometrical results that are summarized in the following Lemma.

\begin{lemma}\label{Lem:Cond}
Let $\varphi = \{(x_1,\tau_1),\dots,(x_n,\tau_n)\}$ be a configuration.
Assume that there exists $(d,D)$ such as $0 < d < D < +\infty$ and
$$
x_i \sim x_j \Rightarrow d \leq |x_i - x_j| \leq D.
$$

Then we have:
\begin{enumerate}
\item the minimum angle in all Delaunay's triangles of $\varphi$ is greater than
$$\beta = \arccos \left(1-\frac{d^2}{2D^2} \right)$$
\item $n(x_i) \leq \frac{2\pi}{\beta}$ where $n(x_i)$ is the number of neighbors of $x_i$ in the configuration $\varphi$,
\item $|{\rm Dir}(x_i \cap x_j)| \leq \frac{D}{\sin(\beta)}$,
\item $|{\rm Dir}(x_i)| \leq \pi \left(\frac{D}{2\sin(\beta)} \right)^2$.
\end{enumerate}
\end{lemma}

\paragraph{Proof of Lemma \ref{Lem:Cond}}\

{\bf 1.} This result is derived from the law of cosines \cite{Abr:72}.

{\bf 2.} comes from item 1. and the fact that each Delaunay's neighbor belongs to two Delaunay's triangles.

{\bf 3.} The radius of a circumcircle to a triangle is equal to $a/\sin(\alpha)$ where $a$ is the length of an edge an $\alpha$ is the opposite angle to the edge. We deduce that the radius of all circumcircle is less than $D/(2\sin (\beta))$. Then we have for all $i \sim j$
$$
|{\rm Dir}(x_i \cap x_j)| < \frac{D}{\sin (\beta)}
$$

{\bf 4.} From 3., we deduce that ${\rm Dir}(x_i)$ is included in the circle of center $x_i$ and of radius $D/ \sin (\beta)$.

\eop

\paragraph{Proof of Lemma \ref{Lem:Stab}}\

This proof can be derived along the same lines as \cite{Ber:99}. Let $\varphi = \{(x_1,\tau_i),\dots,(x_n,\tau_n)\}$ be a (marked) configuration and $(x_{n+1},\tau_{n+1})$ be a (marked) cell. We write
$$
\varphi^{\prime} = \varphi \cup (x_{n+1},\tau_{n+1}).
$$

As Dirichlet cells are associated with a given configuration, we write ${\rm Dir}(x_i)$ (resp. ${\rm Dir}^{\prime}(x_i)$) the Dirichlet cell associated with the cell $x_i$ in the configuration $\varphi$ (resp. $\varphi^{\prime}$). Analogously, ${\rm Dir} (x_i \cap x_j)$ (resp. ${\rm Dir}^{\prime} (x_i \cap x_j)$) characterizes the Dirichlet edge in the configuration $\varphi$ (resp. $\varphi^{\prime}$).

Then, according to the definition of $H$ in Equation \ref{Eq:MyModel} the difference between $H(\varphi^{\prime})$ and $H(\varphi)$ is given by
\begin{eqnarray*}
H(\varphi^{\prime}) - H(\varphi)& = & \sum_{i \sim j} |{\rm Dir}^{\prime}(x_i \cap x_j)|J(\tau_i,\tau_j) +  \lambda \sum_i (|{\rm Dir}^{\prime}(x_i)| - A_{\tau_i})^2 \Gamma(A_{\tau_i}) \\
& & -  \sum_{i \sim j} |{\rm Dir}(x_i \cap x_j)|J(\tau_i,\tau_j) - \lambda \sum_i (|{\rm Dir}(x_i)| - A_{\tau_i})^2 \Gamma(A_{\tau_i})
\end{eqnarray*}

Using geometrical properties of Dirichlet tesselation, the previous expression can be simplified in a sum of four terms.
\begin{eqnarray*}
H(\varphi^{\prime}) - H(\varphi) & = & \sum_{i \sim n+1} |{\rm Dir}^{\prime}(x_i \cap x_{n+1})|J(\tau_i,\tau_{n+1})\\
& & + \lambda (|{\rm Dir}^{\prime}|(x_{n+1}) - A_{\tau_{n+1}})^2\Gamma(A_{\tau_{n+1}})\\
& & + \sum_{{\rm Triangles(i,j,n+1)}} (|{\rm Dir}^{\prime}(x_i \cap x_j)| - |{\rm Dir}(x_i \cap x_j)|) J(\tau_i,\tau_j)\\
& & + \lambda \sum_{i; i \sim n+1} (|{\rm Dir}^{\prime}(x_{i})| - A_{\tau_i})^2\Gamma(A_{\tau_i}) - (|{\rm Dir}(x_{i})| - A_{\tau_i})^2\Gamma(A_{\tau_i})
\end{eqnarray*}

The first term corresponds to the interaction of the additional cell $x_{n+1}$ with its neighbors. The second term is explained by the shape constraint for the additional cell $x_{n+1}$. The third term stand for the difference between the interactions in $\varphi^{\prime}$ and in $\varphi$. The sum runs over all triangles $(i,j,n+1)$ that belong to the Delaunay's configuration. Using geometrical properties about the insertion of a Dirichlet cell, only interactions between two cells that are both in the neighborhood of $x_{n+1}$ are non-zero. The fourth term represents the difference between shape constraints in $\varphi^{\prime}$ and in $\varphi$. According to geometrical properties, the sum runs over all cells in the neighborhood of $x_{n+1}$.

Each term can be controled independently. First, from Lemma \ref{Lem:Cond} item 3, we have $|{\rm Dir}^{\prime}(x_i \cap x_{n+1})| < D/ \sin (\beta)$. Furthermore, Lemma \ref{Lem:Cond} item 2 gives that the number of neighbors of $x_{n+1}$ is less than $N$, which leads to the following result

\begin{equation}\label{Eq:Sup1}
\left| \sum_{i \sim n+1} |{\rm Dir}^{\prime}(x_i \cap x_{n+1})|J(\tau_i,\tau_{n+1}) \right| \leq  N\frac{D}{\sin(\beta)} K_{J}.
\end{equation}

Next, directly from Lemma \ref{Lem:Cond} item 4, we have
\begin{equation}\label{Eq:Sup2}
\left| \lambda (|{\rm Dir}^{\prime}|(x_{n+1}) - A_{\tau_{n+1}})^2\Gamma(A_{\tau_{n+1}}) \right|
\leq  \lambda \pi^2 \left(\frac{D}{2\sin(\beta)} \right)^4
\end{equation}

From Lemma \ref{Lem:Cond} item 3, we have $|{\rm Dir}^{\prime}(x_i \cap x_{n+1})| < D/ \sin (\beta)$. In addition, from Lemma \ref{Lem:Cond} item 2, we extract that the number of pairs in the neighborhood of $x_{n+1}$ is less than $N(N-1)/2$, which leads to the following result
\begin{equation}\label{Eq:Sup3}
\left|\sum_{{\rm Triangles(i,j,n+1)}} (|{\rm Dir}^{\prime}(x_i \cap x_j)| - |{\rm Dir}(x_i \cap x_j)|) J(\tau_i,\tau_j) \right| \leq \frac{N(N-1)}{2} \frac{D}{\sin (\beta)}K_J
\end{equation}

Finally, from Lemma \ref{Lem:Cond} item 2 and \ref{Lem:Cond} item 4 we have
\begin{equation}\label{Eq:Sup4}
\left| \lambda \sum_{i; x_i \sim x_{n+1}} (|{\rm Dir}^{\prime}(x_{i})| - A_{\tau_i})^2\Gamma(A_{\tau_i}) - (|{\rm Dir}(x_{i})| - A_{\tau_i})^2\Gamma(A_{\tau_i}) \right|
\leq \lambda N\pi^2 \left(\frac{D}{2\sin(\beta)} \right)^4
\end{equation}

Combining Equations \ref{Eq:Sup1}-\ref{Eq:Sup4}, we obtain that
$$
\left|H(\varphi^{\prime}) - H(\varphi) \right| \leq  \frac{N(N+1)}{2} \frac{D}{\sin (\beta)} K_J + \lambda (N+1) \pi^2 \left(\frac{D}{2\sin(\beta)} \right)^4
$$

\eop

\newpage
\begin{table}\label{Tab:Estim}
\begin{center}
\begin{tabular}{cp{0.5cm}ccp{0.5cm}cc}
& & \multicolumn{2}{c}{Checkerboarder} & & \multicolumn{2}{c}{Cell Sorting}\\
\hline
& & Mean & Variance & & Mean & Variance\\
\hline
$\theta = 1$ & & 0.98 & 0.70 & & 1.03 & 0.4 \\
$\theta = 3$ & & 3.14 & 0.66 & & 3.01 & 0.51 \\
$\theta = 5$ & & 5.01 & 0.57 & & 4.94 & 0.94 \\
$\theta = 8$ & & 8.20 & 1.07 & & 8.01 & 0.81 \\
$\theta = 10$ & & 10.47 & 1.20 & & 9.80 & 1.00 \\
$\theta = 12$ & & 12.28 & 1.81 & & 12.05 & 1.09 \\
$\theta = 15$ & & 14.58 & 2.22 & & 15.03 & 1.20 \\
$\theta = 20$ & & 20.44 & 3.55 & & 20.08 & 2.98 \\
\end{tabular}
\end{center}
\caption{Mean and Variance of $\hat{\theta}$, maximum of Pseudo-likelihood, for Checkerboard and Cell Sorting simulations. The evaluation was achievied using 50 simulations of each case.}
\end{table}

\newpage

\end{document}